\documentclass{article}

\usepackage{arxiv}

\usepackage[utf8]{inputenc} 
\usepackage[T1]{fontenc}    
\usepackage{hyperref}       
\usepackage{url}            
\usepackage{booktabs}       
\usepackage{amsfonts}       
\usepackage{nicefrac}       
\usepackage{microtype}      
\usepackage{lipsum}
\usepackage{graphicx,verbatim,float,subfigure,graphicx,color,siunitx}
\usepackage{amsmath}
\graphicspath{ {./images/} }

\usepackage{amssymb}
\usepackage{svg}
\usepackage{multirow}  
\usepackage[export]{adjustbox}
\usepackage{soul}
\usepackage{textcomp}
\usepackage{multirow}
\usepackage{enumitem}
\usepackage[sort&compress,numbers]{natbib}

\title{Elucidating the Mechanisms of Damage in Foam Core Sandwich Composites under Impact Loading and Low Temperatures}

\author{
 Alejandra Castellanos and  Pavana Prabhakar$^*$ \\
  Dept. of Civil \& Environmental Engineering \\
  University of Wisconsin-Madison \\
  Madison, WI 53706  \vspace{0.05in} \\
  \texttt{$^*$pavana.prabhakar@wisc.edu}
}

\begin{document}
\maketitle
 

\begin{abstract}
	Recent interest in Arctic exploration has brought new challenges concerning the mechanical behavior of lightweight materials for offshore structures. Exposure to seawater and cold temperatures are known to degrade the mechanical properties of several materials, thus, compromising the safety of personnel and structures. This study aims to investigate the low-velocity impact behavior of woven carbon/vinyl ester sandwich composites with Polyvinyl chloride (PVC) foam core at low temperatures for marine applications. The tests were performed in a drop tower impact system with an in-built environmental chamber. Impact responses, such as the contact force, displacement and absorbed energy, at four impact energies of 7.5 J, 15 J, 30 J, and 60 J were determined at four in-situ temperatures of 25 $^{\circ}$C, 0 $^{\circ}$C, -25 $^{\circ}$C and -50 $^{\circ}$C. Results showed that temperature has a significant influence on the dynamic impact behavior of sandwich composites. The sandwich composites were rendered stiff and brittle as the temperature decreased, which has a detrimental effect on their residual strength and durability. At 7.5 J at all temperatures, the samples experienced matrix cracking, fiber fracture, and delamination at the top face sheet. The samples impacted at 15 J at all temperatures experienced fiber fracture, matrix cracking, and delamination at the top facesheet and localized core crushing/fracture. At 30 J for all the temperatures, the samples exhibited perforation of the top facesheet and penetration into the core. As the temperature decreased, the penetration of the striker into the core increased. At 60 J for all temperatures, the samples experienced perforation of the top facesheet and core, and the back facesheet exhibited varying extent of damage. At -25 $^{\circ}$C and -50 $^{\circ}$CC, the sandwich composite samples were almost completely perforated. At all impact energies, the sandwich composites were rendered stiff and brittle as the temperature decreased, which has a detrimental effect on their residual strength and durability. 
\end{abstract}

\keywords{Foam Core sandwich composites \and Low-velocity impact \and Low temperatures \and Damage mechanisms \and Woven carbon/vinyl ester composites}


\section{Introduction}
An assessment by the U.S. Geological Survey (USGS) of the Arctic circle has revealed that approximately 22\% of the world's undiscovered oil and natural gas resources are located in this region \cite{USGS2008}. Due to ice melting in the Arctic, new passages (Northwest and Northeast) have been created, which can become viable transportation routes \cite{Arctic2009} in the future. Therefore, there has been a high interest in Arctic explorations resulting in the investigation of the mechanical behavior and damage mechanisms of current materials when exposed to low temperatures. Offshore structures, such as ship vessels, are typically subjected to adverse environments such as sea water, wave impacts and extremely low temperatures. These environments can cause surface alterations, internal damage, and degradation of the chemical and mechanical properties that may ultimately compromise structural safety. Therefore, the materials used in these structures must be able to withstand harsh environmental conditions in addition to mechanical loads.

Sandwich composites have become an attractive option for marine applications due to their lightweight, high stiffness and high strength to weight ratio \cite{Krzyzak2016}. They consist of two thin but very stiff facesheets that sandwich a thick lightweight core between them. The facesheets carry transverse loads or bending moments while the core carries transverse shear loads \cite{Jones2007}. The separation of the facesheets increases the moment of inertia of the panels with little increase in weight, which results in an efficient structure that can resist bending and buckling loads \cite{Petras1998}. Despite several advantages of sandwich composites, a major drawback is their low resistance to impact damage. Dynamic impact on structures can occur under different scenarios, for example, tool drop during maintenance and repair, wave slamming, iceberg collisions, bird or hail strikes \cite{julias2014,Dempsey2000,Zhu1996}.

Low-velocity impacts typically occur at velocities below 10 m/s \cite{sjoblom1988}, which may produce barely visible damage (BVD) on composite surfaces (facesheets), but with potential significant internal damage. This is deemed very dangerous, as BVD could result in catastrophic failure of the structure without warning. Therefore, low-velocity impact studies on sandwich composites is critical for material certification and establishing allowable for structural design. In particular, their response to dynamic impact loading at low temperatures is of main interest in this paper in light of arctic applications. 

There is a wealth of research studies available that have focused on the dynamic impact behavior of foam core sandwich composites  \cite{Jover2014,  Tagarielli2007, Torre2000, Wang2015, Ozdemir2015, Schubel2004, Loganathan2015, Park2008, Gupta2012, Gibson1997, Erickson2005, Salehi2007}. Most of these prior studies have focused on low-velocity impact loading at room temperature (25 $^{\circ}$C). Ozdemir et al. \cite{Ozdemir2015} investigated the effect of core material and thickness (5, 10 and 15 mm) of unidirectional E-glass/epoxy sandwich composite panel with PVC and poly(ethylene terephthalate) (PET) core with impact energies ranging from 10 J to 70 J at room temperature. They reported that the specimens absorbed more energy as the core thickness increased. Also, PVC cores had higher bending stiffness than PET cores. Schubel et al. \cite{Schubel2004} studied the quasi-static and low-velocity (1.6 to 5 m/s) impact behavior of woven carbon/epoxy sandwich composites with PVC foam core. They concluded that the low-velocity impact response of plates could be characterized as quasi-static based on the load-strain response and damage evaluation. Loganathan et al. \cite{Loganathan2015} explored the effect of core thickness (10 and 14 mm) and density (70, 100 and 200 kg/$m^3$) of bi-directional woven E-glass/epoxy sandwich composite with polyurethane foam core under three different impact velocities (1.401, 2.426 and 3.123 m/s). They observed that the samples absorbed more energy with increasing core density and thickness. Park et al. \cite{Park2008} evaluated the impact damage resistance of unidirectional carbon/epoxy and glass epoxy sandwich composites with Nomex\textsuperscript{\textregistered} honeycomb core at room temperature. They concluded that the damage resistance of sandwich structures is dependent on the facesheet material and core thickness.

Despite an extensive amount of research reported at room temperature, relative less research has been conducted at low temperatures. Gupta et al. \cite{Gupta2012} studied the blast performance of E-glass/ vinyl ester sandwich composites with Corecell\textsuperscript{\texttrademark} M100 foam core at three different temperatures (22 $^{\circ}$C, 80 $^{\circ}$C and -40 $^{\circ}$C) and average strain rate varied from 1600/s (-40 $^{\circ}$C) to 2000/s (100 $^{\circ}$C). They reported that the plateau stress reduced with increase in temperature from -40 $^{\circ}$C to 100 $^{\circ}$C, which was attributed to thermal softening in the core that resulted in the collapse of cells under compression due to shock wave loading. In addition, the glass facesheets showed an increase in damage (delamination and fiber breakage) as the temperature increased from -40 $^{\circ}$C to 100 $^{\circ}$C. This was attributed to the softening of matrix in the facesheet, which led to the decrease in the compressive modulus and compressive strength of the composite. Erickson et al. \cite{Erickson2005} investigated the low-velocity impact behavior of woven E-glass/epoxy sandwich composites with filled and non-filled honeycomb core subjected to three different temperatures (-25, 25 and 75 $^{\circ}$C) under three impact energies (12, 60 and 150 J), and concluded that the maximum impact force decreased and the energy absorption increased with increasing temperature (-25 to 75 $^{\circ}$C). Salehi-Khojin et al. \cite{Salehi2007} explored the effects of temperature (-50 $^{\circ}$C to 120 $^{\circ}$C) on woven carbon fiber/epoxy, Kevlar\textsuperscript{\textregistered}/epoxy and hybrid Kevlar\textsuperscript{\textregistered}-carbon/epoxy sandwich composites with polyurethane foam filled with Kraft paper honeycomb core subjected to impact energies of 15, 25 and 45 J. They observed that the largest area of damage and fiber breakage occurred at -50 $^{\circ}$C, which decreased with increasing temperature. There are more studies on the effects of temperature on fiber reinforced laminates  \cite{Icten2015,Ibekwe2007,Lopez-Puente2002,Im2001}. Castellanos et al. \cite{Castellanos2018,Castellanos2018b} studied the single and repeated impact behavior of woven carbon/vinyl ester composites at room and arctic (-50 $^{\circ}$C) temperatures subjected to four impact energies (20, 25, 30 and 35 J). They reported that the damage mechanisms shifted from predominantly matrix cracking to fiber fracture when the in-situ temperature was changed from room to arctic. This shift was more noticeable at lower impact energies.

Understanding the behavior of sandwich composites at low temperatures is critical for the safety of structures and personnel. In the current study, the mechanical response and damage mechanisms of woven carbon/vinyl ester laminated sandwich composites with Polyvinyl chloride (PVC) foam core subjected to a range of low-velocity impact loading (1.22, 2, 2.45 and 3.46 m/s) at four different temperatures (25, 0, -25 and -50 $^{\circ}$C) are investigated. The variations in impact response in terms of force, displacement, energy and damage mechanisms are studied in detail and presented here. In addition, a detailed visual damage investigation is conducted using micro-Computed tomography to identify key damage mechanisms that can potentially serve as a guide for developing relevant repair and reinforcing techniques for foam core sandwich composites.

\section{Methods}
\subsection{Panel Fabrication}
\begin{figure}[!htb]
	\centering
	\subfigure[]{
		\includegraphics[width=7.7cm]{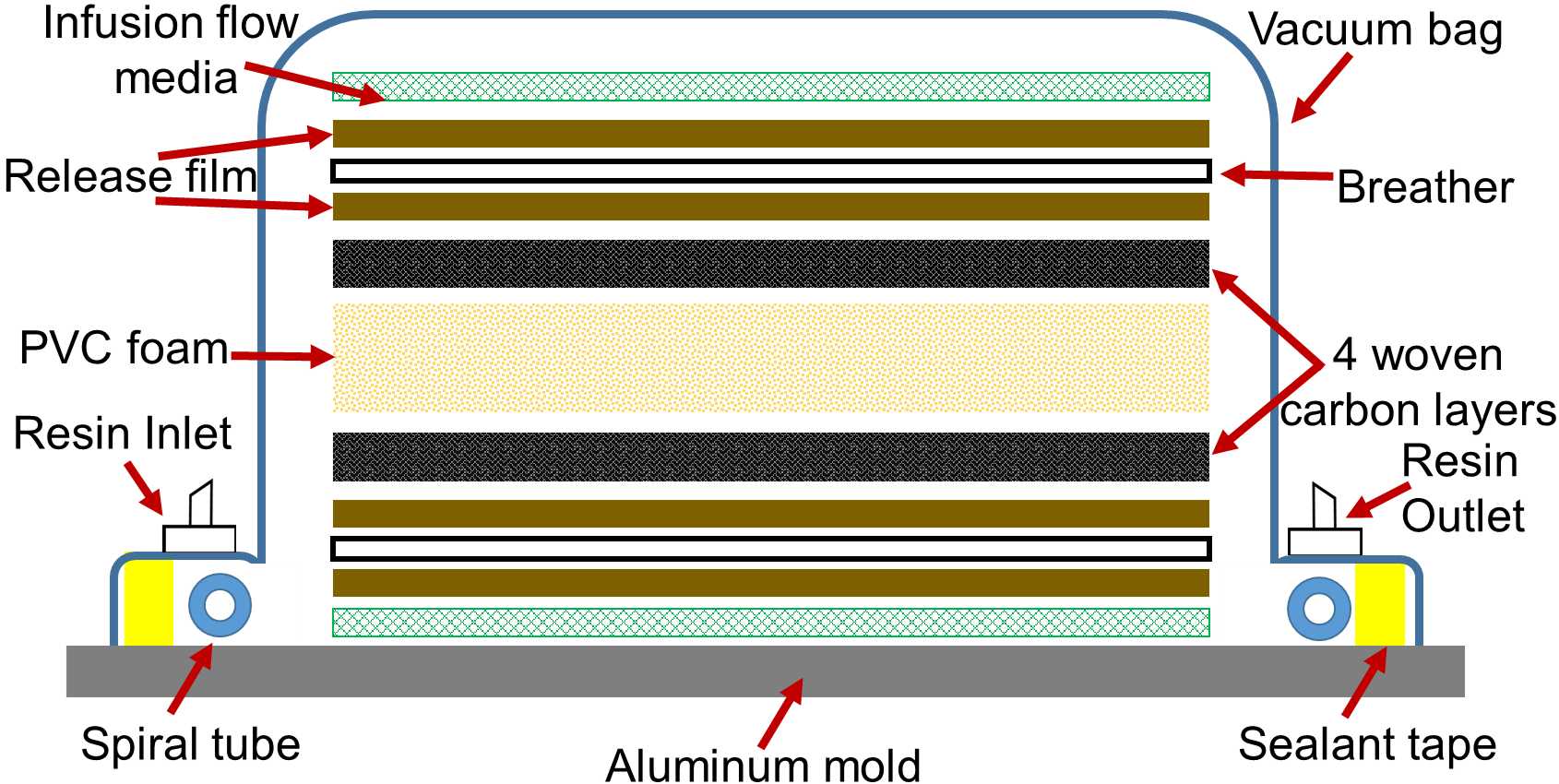}
		\label{VARTM}
	}
	\subfigure[]{
		\includegraphics[width=6.7cm]{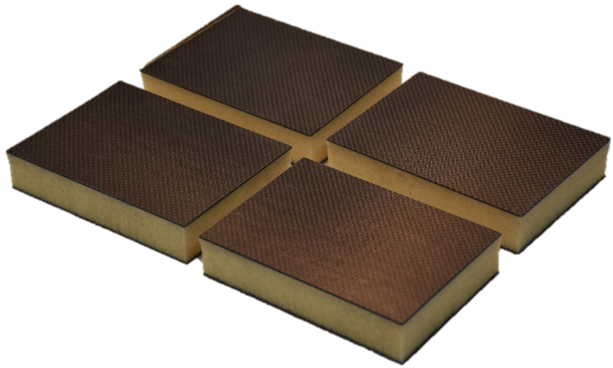}
		\label{waterjet}
	}
	\caption[Optional caption for list of figures]{(a) VARTM process configuration; (b) Samples obtained from each sandwich composite panel}
\end{figure}

Woven carbon fiber/vinyl ester sandwich composites with PVC core were manufactured by vacuum assisted resin transfer molding (VARTM) process \cite{VARTM} as seen in Figure~\ref{VARTM}. Woven plain weave carbon fabric and vinyl ester resin were purchased from Fibre Glast (www.fibreglast.com), and the H100 PVC foam was purchased from Aircraft Spruce (www.aircraftspruce.com). Vinyl ester was considered as the resin due to superior UV resistance and low water absorption as compared to polyester resins \cite{Sobrinho2009,Signor2003}. PVC was considered due to their low cost, low density, fire retardancy and high insulation and damping properties \cite{Jamel2015}. The mechanical properties of the constituent materials are given in Table~\ref{t:mechprop}.

\begin{table}[!htb]
	\small\sf\centering
	\begin{center}
		\centering
		\caption{Mechanical properties of the sandwich composite constituent materials\label{T1}}
			\label{t:mechprop}
		\begin{tabular}{ccccc}
			\toprule
			Mechanical &    Woven & Vinyl & \multicolumn{2}{c}{H100 foam\footnote[3]{normal footnote.}}\\
			properties & carbon& ester & In & Out of\\
			& fiber\footnote[1]{normal footnote.} & resin\footnote[2]{normal footnote.} & Plane & Plane\\
           			\midrule
			Tensile& \multirow{2}{*}{2275-2406}& \multirow{2}{*}{3700} & \multirow{2}{*}{111} &  \multirow{2}{*}{126}  \\
			Modulus (MPa) & &  &\\
			Tensile & \multirow{2}{*}{4200-4400}& \multirow{2}{*}{82.7} & \multirow{2}{*}{3} &  \multirow{2}{*}{3.3}  \\
			Strength (MPa) &  &  & \\
			Density (kg/$m^3$) & 1750-2000& 1.800 & 100 &100  \\
			Nominal & \multirow{2}{*}{0.3048}& \multirow{2}{*}{-} & \multirow{2}{*}{25.4} &  \multirow{2}{*}{25.4}  \\
			thickness (mm) &  &  & \\
			\bottomrule
		\end{tabular}\\[10pt]
\footnote[1]~\cite{Fibrewcf2010}
\footnote[2]~\cite{fvinyl2010}
\footnote[3]~\cite{Viana2002}
	\end{center}
\end{table}

The fabrication of sandwich composite panels consisted of a PVC foam sandwiched between two facesheets with [$(0/90)_{4}$/core/$(0/90)_{4}$] stacking sequence as per the instructions in ASTM D7766/D7766M-16 \cite{ASTMD7766}. Here, $(0/90)_{4}$ implies 4 layers of plain weave carbon fabrics. This arrangement was placed between two layers of flow-media, two layers of breather and four layers of nylon peel ply. All the layers were cut to dimensions of 305 mm x 254 mm. This complete arrangement of fabrics was placed over an aluminum mold, then wrapped with two Stretchlon 800 bagging film and sealed with vacuum-sealant tape, ensuring spaces for both inlet and outlet connectors. The first vacuum bag assisted with resin infusion, and the second vacuum bag applied continuous pressure during the curing process to obtain superior surface finish on the panels. A mixture of vinyl ester resin and Methyl Ethyl Ketone Peroxide (MEKP) hardener was used for resin infusion to reinforce the dry woven fabrics and to bond the facesheets to the core. The resin was catalyzed with 1.25\% MEKP by weight and mixed thoroughly as recommended by the manufacturer. The resin/hardener mixture was placed in a desiccator to remove any air bubbles from the mixture. The outlet was then connected to a vacuum pump until the vacuum bag achieved a pressure of approximately 80 MPa. The inlet of the vacuum bag was then submerged in the resin/hardener mixture to infuse the resin through the sandwich panel. Upon completion of the resin transfer process, the panel was cured at room temperature for 24 hours under constant pressure of $\approx$80 MPa by the second vacuum bag. 

A total of 16 panels of 305 mm length x 254 mm width were manufactured. Four samples (shown in Figure~\ref{waterjet}) were water jet cut from each panel. Impact test samples were randomly chosen from the set of all the samples fabricated to distribute any manufacturing induced effects on the impact response. Four samples were selected for testing for each combination of temperature and impact energy.

\subsection{Impact Tests}
\begin{figure}[!htb]
	\begin{center}
		\includegraphics[width=8.5cm]{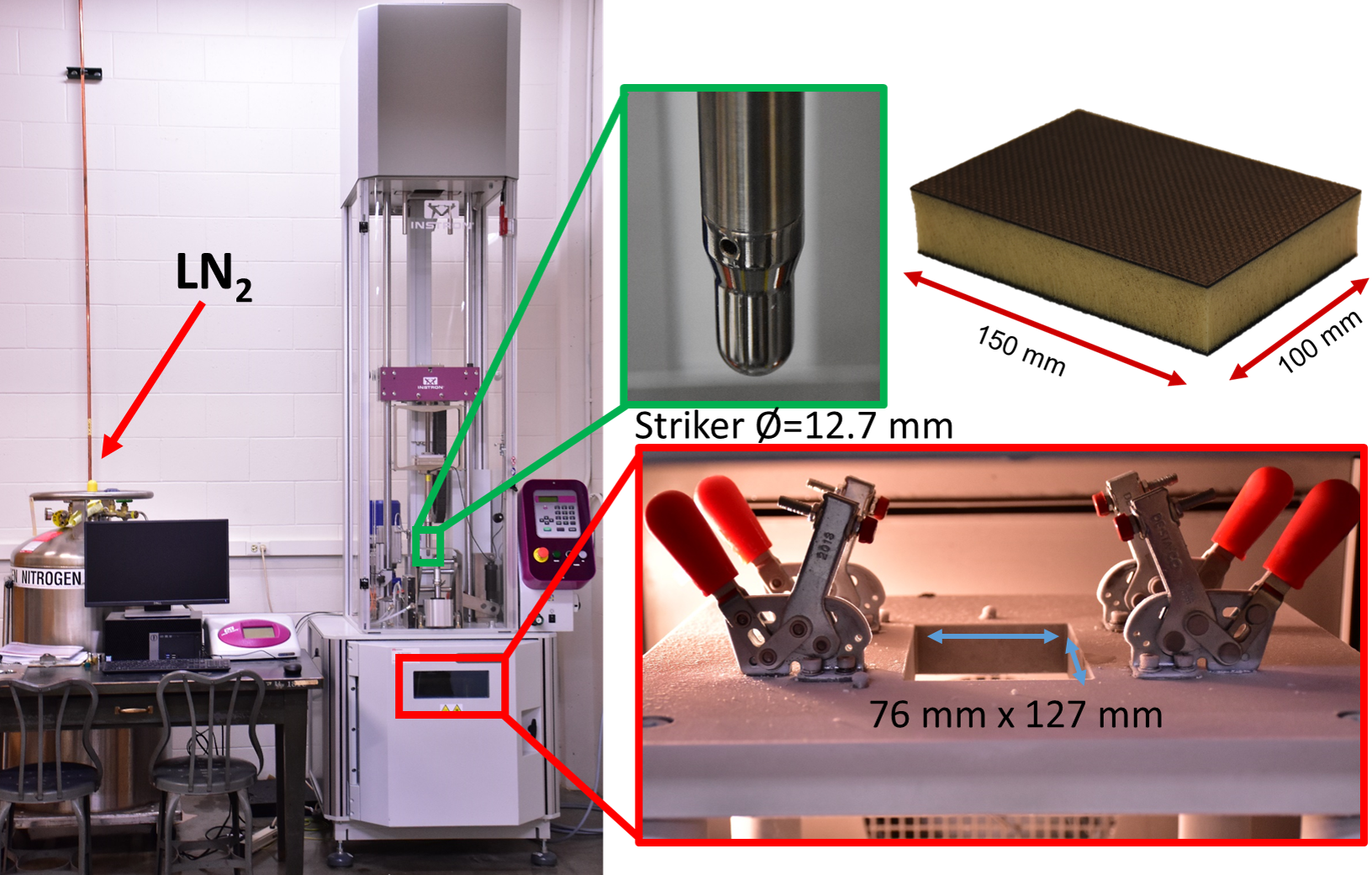}
		\caption{CEAST 9350 Drop Tower Impact System}\label{CEAST}
	\end{center}
\end{figure}

Drop-weight impact tests were performed using a CEAST 9350 Drop Tower Impact System with a load cell capacity of 22.4 kN and an in-built environmental chamber as shown on Figure~\ref{CEAST}. A metal fixture with a rectangular opening of 76 mm x 127 mm and toggle clamps were used. These clamps apply pressure in the vicinity of the four corners of the samples to prevent their motion during an impact event. The rectangular sandwich samples had dimensions of 150 mm length x 100 mm width (refer to Figure~\ref{CEAST}) with an average thickness of 27.43$\pm$0.16 mm. The core thickness was 25.4 mm, and each facesheet had an approximate thickness of 1 mm.  A hemispherical striker with a fixed mass of 10 kg and a diameter of 12.7 mm was used to impact the samples at the center of the top facesheet in the out-of-plane direction \cite{ASTMimpact}. The impact energies chosen were 7.5 J, 15 J, 30 J, and 60 J, which corresponded to impact velocities of 1.22 m/s, 2 m/s, 2.45 m/s and 3.46 m/s, respectively. All these tests were conducted at low-velocities, that is, below 10 m/s \cite{Yang2015,cantwell1990}. The kinetic energies corresponding to these velocities were calculated based on the mass of the striker and the impact velocities using the equation $E_{k}=\frac{1}{2}mv^{2}=mgh$. Here, $E_{k}$ is the impact energy or kinetic energy, $v$ is the impact velocity, and $m$ is the mass of the impacting striker, $h$ is the height of the striker measured from the surface of a sample in the impact drop tower, and $g$ is the gravitational acceleration. 

In order to investigate the influence of temperature on the impact response of sandwich composites, four in-situ temperatures were selected: 25 $^{\circ}$C, 0 $^{\circ}$C, -25 $^{\circ}$C and -50 $^{\circ}$C. Robinson et al. \cite{Robinson1992} investigated the influence of impactor mass and velocity on the low velocity impact response of woven carbon and glass fiber reinforced laminates for impact energies ranging from 0.25 J to 12 J. The impactor mass was varied from 1.15 kg to 2.10 kg and the velocity was adjusted to obtain the desired impact energies. They reported that the extent of impact damage predominantly depended on the magnitude of the impact energies and less on the mass or velocity individually. Based on this study, the impactor mass was held constant and the impact energy was varied for the prescribed velocities in the current paper.

Force-time, displacement-time and energy-time responses were recorded by the data acquisition system ``CEAST DAS 8000 Junior'' of the impact machine for each test. Four samples were impacted for each combination of impact energy and temperature. The samples impacted at temperatures different than 25 $^{\circ}$C were placed in a temperature controlled environmental chamber, which was connected to the CEAST 9350 Drop Tower Impact System. Prior to every impact test, the samples were conditioned in the environmental chamber for 20 minutes with Liquid Nitrogen $(LN_2)$ to reach the desired temperature throughout the sample.

\subsection{Micro-CT Scanning}
Sandwich composite samples were examined under a micro-computed tomography (micro-CT) scanner, a non-destructive technique (NDT), after testing. The goal was to evaluate the internal damage incurred at different temperatures and impact energies. The samples were scanned within a Zeiss Metrotom OS 800 scanner at UW-Madison. For each sample, 1500 projections were made by rotating them until a complete revolution was obtained. The X-ray tube voltage and current were set to 80 keV and 120 $\mu$A, respectively. Reconstruction of the 3D virtual object was done with METROTOM OS software, and were further analyzed with VG Studio Max 22 software. A flat correction was applied for each scan. 

\subsection{Quasi-static tests to establish the effects of temperature on matrix and fibers}
To further examine the influence of low temperature on sandwich composites, compressions tests were performed on pure vinyl ester samples and tension tests were performed on woven carbon/vinyl ester samples at 25 $^{\circ}$C (RT) and -50 $^{\circ}$C (AT). For the compression tests, three pure vinyl ester samples were tested in-situ at 25 $^{\circ}$C and -50 $^{\circ}$C under flat-wise compressive loading. Cylindrical samples with a diameter of 25.5 mm and a height of 50.8 mm were tested according to ASTM D695 \cite{ASTMcompression}. The tests were performed using an ADMET eXpert 1654 testing system with a crosshead displacement rate of 1.3 mm/min. For the tension tests, five woven carbon/vinyl ester samples were tested in-situ at 25 $^{\circ}$C and -50 $^{\circ}$C. Rectangular samples with a width of 15 mm, length of 250 mm and thickness of 1 mm were tested according to ASTM D3039 \cite{ASTMD3039}. These tests were also performed using the ADMET eXpert 1654 testing system mentioned above with a crosshead displacement rate of 2 mm/min.

\section{Results and Discussion}
Prior to discussing the results of the dynamic impact tests, quasi-static test results of pure vinyl ester and woven carbon/vinyl ester laminates are discussed to provide insight on the influence of temperature on the matrix and fibers independently.

\subsection{Influence of temperature on the matrix and fibers under quasi-static loading}\label{tempeffect}
Prior research by Dutta \cite{Dutta1994} on the compressive response of glass fiber-reinforced polymer composites at the U.S. Army Cold Regions Research and Engineering Laboratory (CRREL) reported that the quasi-static strength and stiffness increased at low temperatures. However, they were also rendered brittle and manifested an increase in cracking due to higher residual thermal stresses caused by a mismatch in the coefficient of thermal expansion (CTE) between fibers and matrix. Similar behaviors were recorded in the current study as discussed next.

\begin{figure}[!htb]
	\centering
	\subfigure[]{
		\includegraphics[height=6.cm]{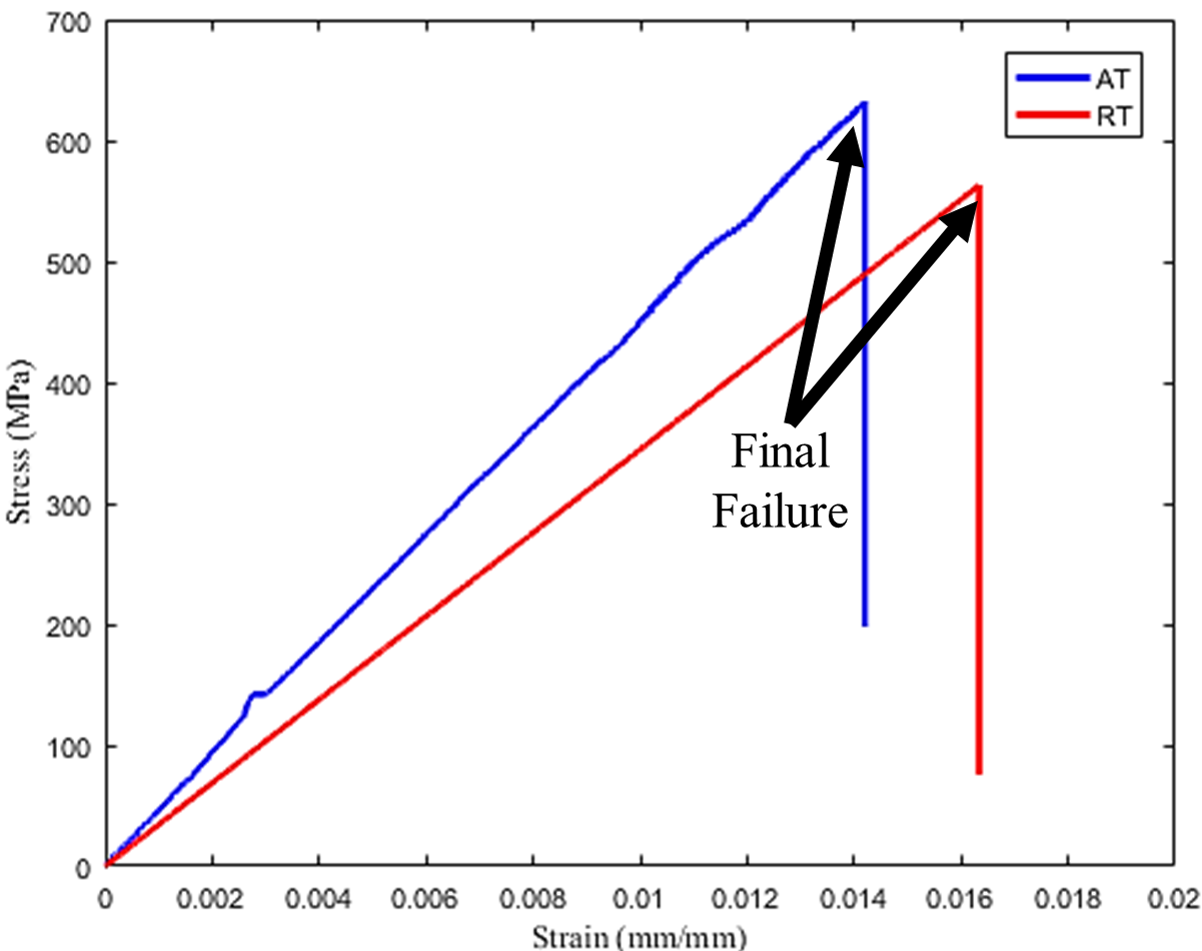}
		\label{tensile}
	}
	\subfigure[]{
		\includegraphics[height=6.cm]{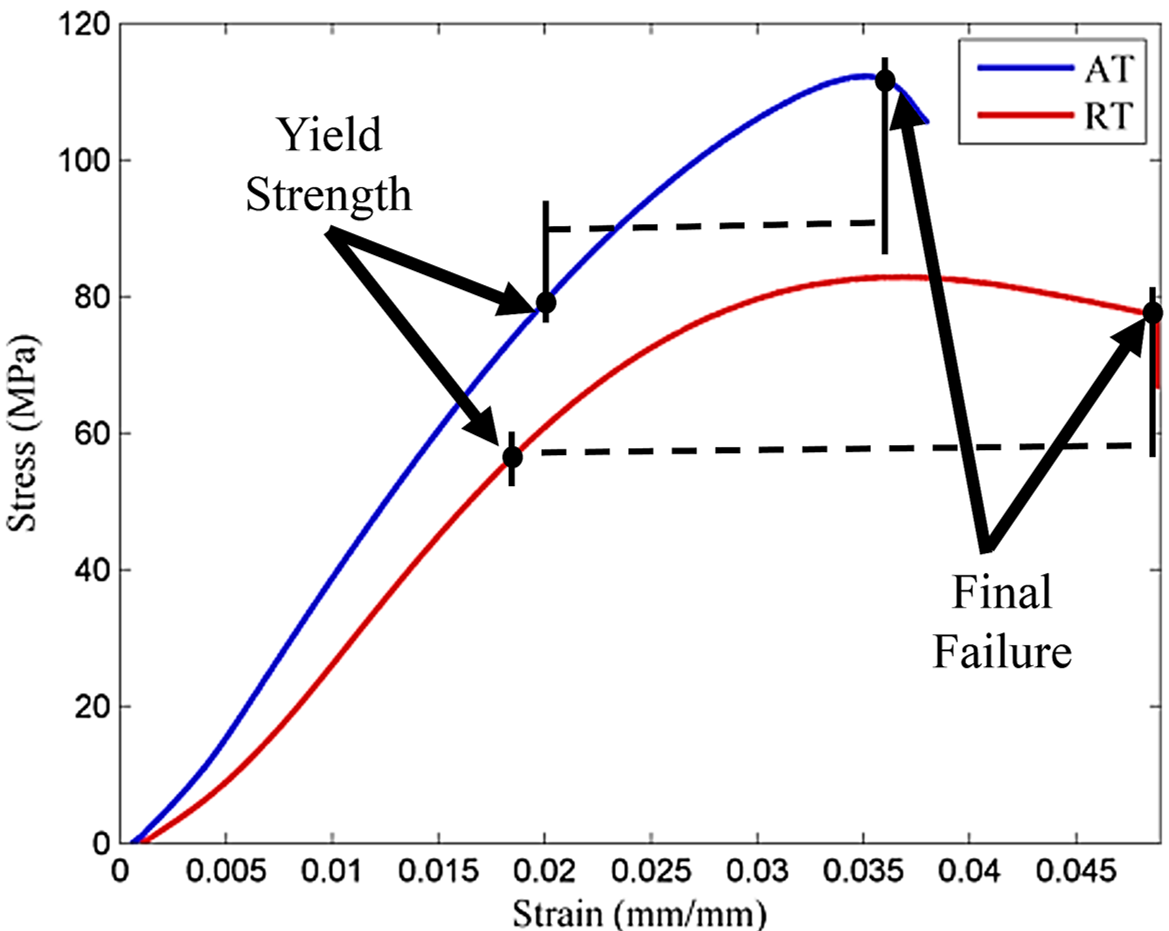}
		\label{compressive}	
	}
	\caption[Optional caption for list of figures]{(a) Typical tension stress-strain plots of woven carbon/vinyl ester at RT and AT, and (b) Typical compressive stress-strain plots of vinyl ester at RT and AT}
\end{figure}

Tension tests on woven/carbon vinyl ester composite samples were conducted in the current study at 25 $^{\circ}$C (RT) and -50 $^{\circ}$C (AT). It was observed that the elastic modulus and ultimate tensile strength increased by approximately 15\% and 11\%, respectively, while the strain to failure reduced approximately by 14\% when the temperature decreased from 25 $^{\circ}$C to -50 $^{\circ}$C. Representative tensile stress-strain responses at RT and AT are shown in Figure~\ref{tensile}, which show a reduction in the ductility and increase in the brittleness at AT \cite{Hartwig1984}. Kim et al. \cite{Kim2007} attributed the increase in brittleness of the laminate under tension to the fibers. As the temperature decreased from 25 $^{\circ}$C to -50 $^{\circ}$C, the brittleness of the fibers increased.

Compression tests on pure vinyl ester were conducted in RT and AT. Representative compressive stress-strain responses of samples tested at 25 $^{\circ}$C and -50 $^{\circ}$C are shown in Figure ~\ref{compressive}. When the temperature decreased from RT to AT, the yield strength (here denoted as the stress recorded at $\approx$0.02 strain), ultimate strength and elastic modulus increased by approximately 55, 49 and 28\%, respectively. However, the strain to failure reduced by approximately 30\% with the decreased of temperature. This implies that the apparent elastic modulus and compressive strength of the resin increased at low temperatures. On the other hand, the resin also became more brittle at low temperatures and therefore was unable to withstand large deformations as compared to the samples tested at 25 $^{\circ}$C. 

\subsection{Dynamic Impact Tests}
Contact force-time, energy-time, and deformation-time responses were recorded by the data acquisition system of the impact machine for each combination of impact energy (7.5 J, 15 J, 30 J and 60 J) and temperature (25 $^{\circ}$C, 0 $^{\circ}$C, -25 $^{\circ}$C and -50 $^{\circ}$C) tested. The response of the panels in terms of visual damage and damage mechanisms were established, and are discussed in detail in the following sections. 

\subsubsection{Contact Force-Displacement Response}
\begin{figure}[!htb]
	\centering
	\subfigure[]{
		\includegraphics[height=5.0cm]{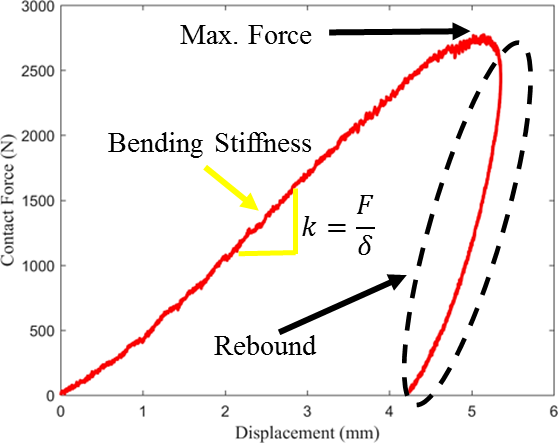}
		\label{rebounding}
	}
	\subfigure[]{
		\includegraphics[height=5.0cm]{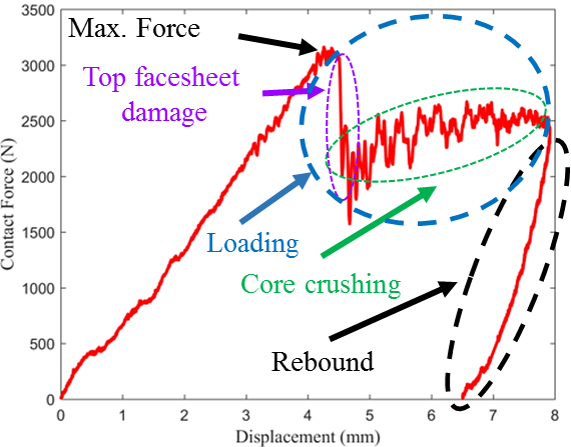}
		\label{partial}
	}
	\subfigure[]{
		\includegraphics[height=5.0cm]{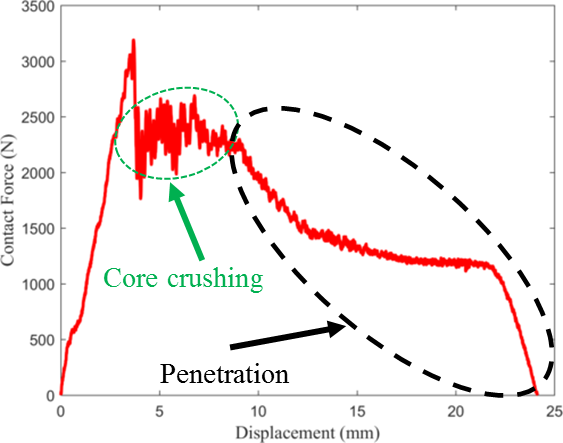}
		\label{penetration}
	}
	\subfigure[]{
		\includegraphics[height=5.0cm]{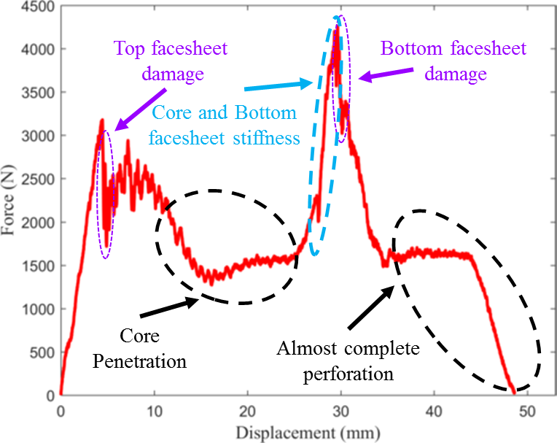}
		\label{perforation}
	}
	
	\caption[Optional caption for list of figures]{Typical load deflection responses of foam core sandwich composites under low-velocity impact loading: (a) Rebounding, (b) Partial loading and rebound, (c) Partial Perforation and (d) Complete Perforation }\label{characteristic}
\end{figure}

~\\
Contact of the impact striker with the impacted face of a sandwich composite sample (top facesheet) generates the contact force and deflection during an impact test, which is recorded by the data acquisition system of the impact machine. Typically, contact force-deflection responses have an initial linear ascending region from which the initial bending stiffness \cite{abrate1998} is determined. These graphs were obtained from the impact tests performed in the current study. Four characteristic dynamic impact responses are shown in Figure~\ref{characteristic}, where the post-peak regimes upon reaching the maximum contact force have very different behaviors. There can be three possibilities: rebounding, penetration, and perforation by the striker. In Figure~\ref{rebounding}, the striker rebounds from the sample upon reaching a maximum contact force. A rebounding case is characterized by a gradual decrease in the displacement while the contact force diminishes to zero. Figure~\ref{partial} shows a typical response of a descending regime that combines loading and unloading (rebounding). Here, there is a sudden drop in load after the maximum contact force is reached, which is typically due to damage in the top facesheet (impacted surface), such as fiber fracture, delamination or matrix cracks. This is followed by a small increase in load as the displacement increases, which is due to the crushing of the core by the striker. This is expected as energy dissipation occurs through plastic deformation of the core and damage in the facesheet \cite{Massabo2012}. Similar behavior has been observed previously by Daniel \cite{Daniel2010b}. Finally, the striker rebounds without damaging the bottom facesheet. Figure~\ref{penetration} shows a representative case of striker penetration, where the striker perforates the top facesheet and the core of the sandwich composites, with no damage in the bottom facesheet. The displacement continues to increase with reduction in contact force, which implies that there is no rebound. Figure~\ref{perforation} shows a representative case of perforation of the sandwich composite by the striker. This graph is characterized by two peak forces. The first peak and its post-peak response represents the perforation of the top facesheet and the core by the striker. Beyond that, the contact force increases, resulting in a second peak, which is larger than the first peak. This is attributed to the stiffness of the bottom facesheet which is subjected to local in-plane tensile loading. If the bottom facesheet is damaged, a sudden drop in contact force follows the second peak, which is typically due to fiber fracture. This is followed by an increase in displacement with reduction in contact force, which implies complete perforation of the sandwich composite.

\subsubsection{Influence of temperature on the contact force - displacement responses}
\begin{figure}[!htb]
	\centering
	\subfigure[]{
		\includegraphics[height=5.0cm]{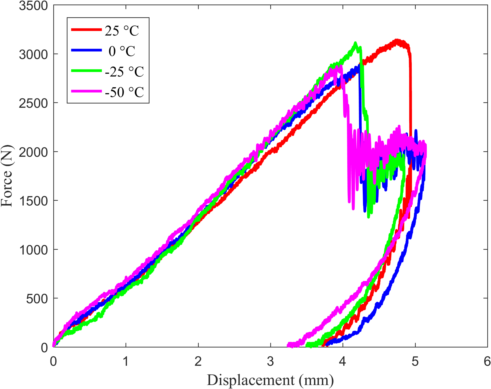}
		\label{7.5}
	}
	\subfigure[]{
		\includegraphics[height=5.0cm]{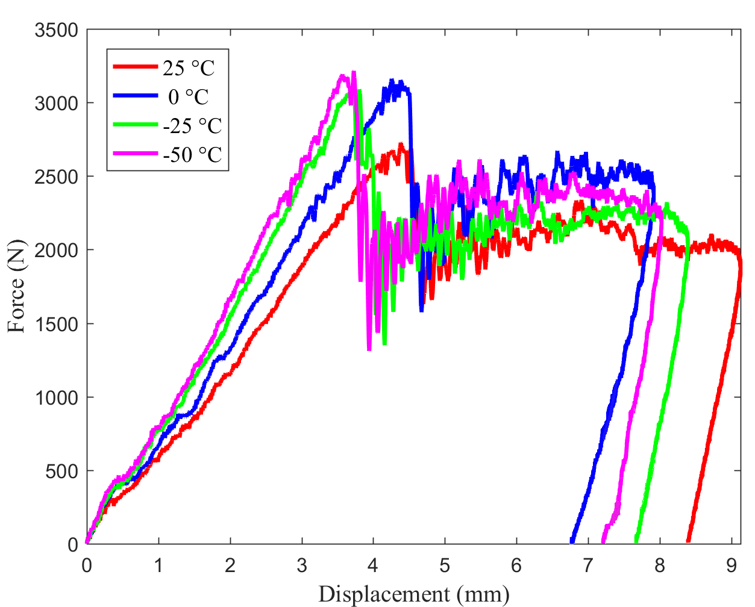}
		\label{15}
	}
	\subfigure[]{
		\includegraphics[height=5.0cm]{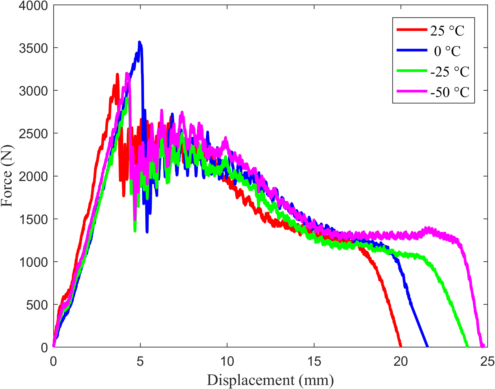}
		\label{30}
	}
	\subfigure[]{
		\includegraphics[height=5.0cm]{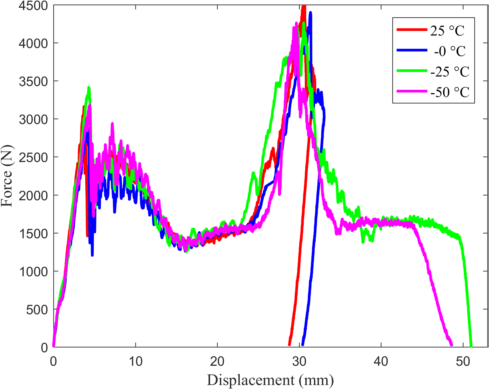}
		\label{60}
	}

	\caption[Optional caption for list of figures]{Representative contact force-displacement responses for all temperatures under impact energies of: (a)7.5 J, (b) 15 J, (c) 30 J and (d) 60 J.}
\end{figure}

~\\
In this section, the contact force-displacement responses were used to characterize the damage caused by an impact event at four different impact energies (7.5 J, 15 J, 30 J and 60 J) and at four distinct temperatures (25 $^{\circ}$C, 0 $^{\circ}$C, -25 $^{\circ}$C and -50 $^{\circ}$C). Figure~\ref{7.5}, Figure~\ref{15}, Figure~\ref{30} and Figure~\ref{60} show the representative force-displacement responses for each case.

Figure~\ref{7.5} corresponds to an impact energy of 7.5 J. A drop in the contact force is observed upon reaching the maximum contact force at 25 $^{\circ}$C. This indicates damage in the top facesheet, upon which the striker rebounds from the sample. However, at 0 $^{\circ}$C, -25 $^{\circ}$C and -50 $^{\circ}$C responses, damage in the top facesheet is accompanied with core crushing. This is indicated by the small increase in the contact force after the first vertical drop. Finally, the striker rebounds from the specimens without further penetration into the samples. This can be seen by the gradual decrease in the displacement while the contact force diminished to zero. Figure~\ref{15} corresponds to an impact energy of 15 J, where all the graphs manifest a sudden drop after the maximum contact force is reached, which implies that  the top facesheet experienced damage along with a noticeable amount of indentation on the impacted surface. After that, the striker rebounds from the specimens. The indentation is manifested by an increase in residual displacement of the striker after rebound. For this impact energy, only crushing of the core and perforation of the top facesheet by the striker is observed. Figure~\ref{30} corresponds to an impact energy of 30 J, where the striker perforates the top facesheet (drop in load after the maximum force is reached) and penetrates through the foam core at all temperatures. The core penetration by the striker is indicated by a decrease in the contact force with increasing displacement after the drop in load of the maximum contact force. 

Figure~\ref{60} corresponds to an impact energy of 60 J. All the responses have two peaks for contact force. The first peak and post-peak response correspond to the perforation of the top facesheet and core penetration by the striker, respectively. The second peak corresponds to damage of the bottom facesheet. The second peak is higher than the first peak due to higher strength of the bottom facesheet under tension \cite{Al-Shamary2016} as compared to the compressive strength of the top facesheet. The graphs at 25 $^{\circ}$C and 0 $^{\circ}$C have the same profile. After the second peak, the displacement of the striker diminishes to zero as the contact force reduces to zero. This implies that the striker reaches the bottom facesheet, stresses it and then rebounds from the bottom facesheet \cite{Atas2010}. In contrast, at -25 $^{\circ}$C and -50 $^{\circ}$C, the striker almost entirely perforates the samples and gets lodged in the bottom facesheet with no rebound. This can be seen after the second peak, where the contact force diminishes to zero as the displacement increases. These damage modes are described in more detail in the ``Damage Mechanisms'' section. 

\begin{figure}[!htb]
	\centering
	\subfigure[]{
		\includegraphics[height=4.5cm]{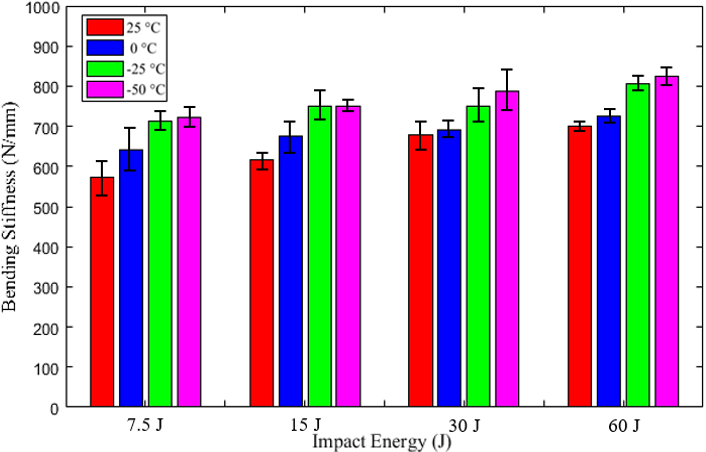}
		\label{bending}
	}
	\subfigure[]{
		\includegraphics[height=4.5cm]{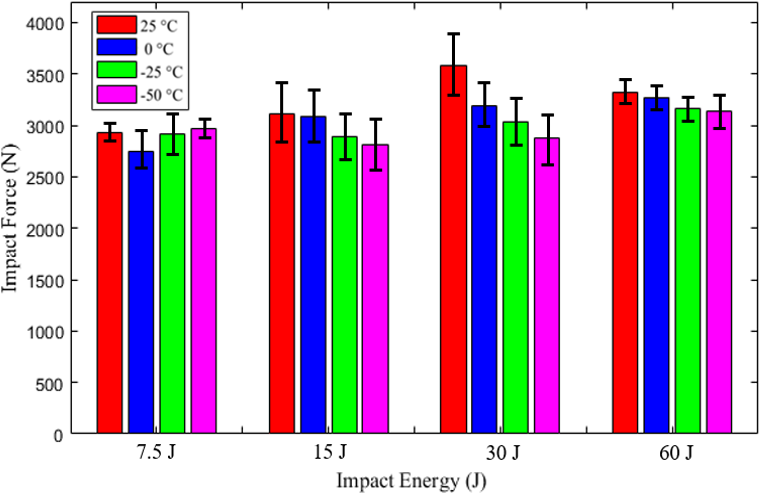}
		\label{maxforce}
	}
	
	\caption[Optional caption for list of figures]{(a) Average bending stiffness and (b) Average first maximum contact force with increasing impact energy and four different temperatures (25 $^{\circ}$C, 0 $^{\circ}$C, -25 $^{\circ}$C and -50 $^{\circ}$C).}
\end{figure}

Figure~\ref{bending} shows the average bending stiffness values that were calculated using the initial ascending regime of the contact force - displacement responses, which corresponds to the top facesheet. At all impact energies, the bending stiffness increases with reducing temperature, which is expected based on the description given in subsection ``Influence of temperature on the contact force-displacement responses''. The first maximum contact force values (first peak on a contact force-displacement graph) is shown in Figure~\ref{maxforce}. For impact energies of 15 J, 30 J and 60 J, the contact force has a decreasing trend as the temperature decreases from 25 $^{\circ}$C and -50 $^{\circ}$C. This is attributed to the top facesheet failure due to embrittlement of sandwich composites at low temperatures. At an impact energy of 7.5 J, the peak contact forces oscillate with changing temperature, as the top facesheet is minimally damaged within the prescribed temperature range (25 $^{\circ}$C and -50 $^{\circ}$C).

\subsection{Absorbed Energy}
\begin{figure}[]
	\centering
	\subfigure[]{
		\includegraphics[height=5.0cm]{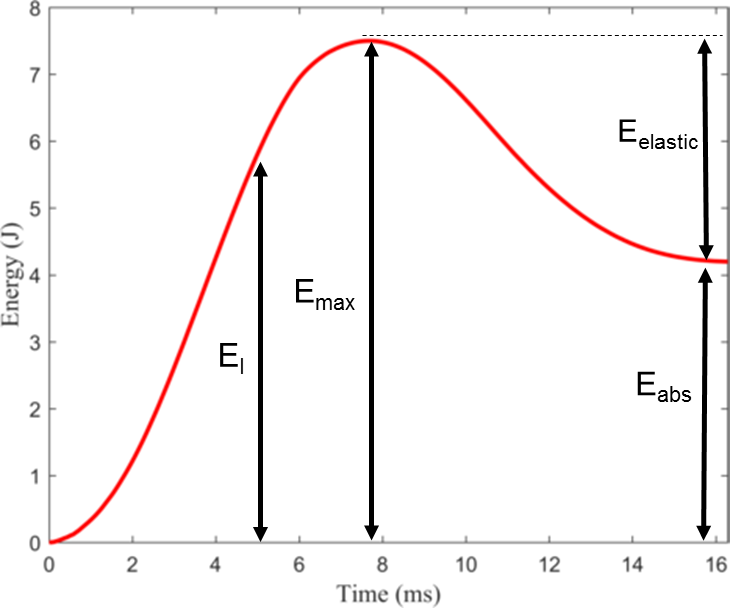}
		\label{7525}
	}
	\subfigure[]{
		\includegraphics[height=5.0cm]{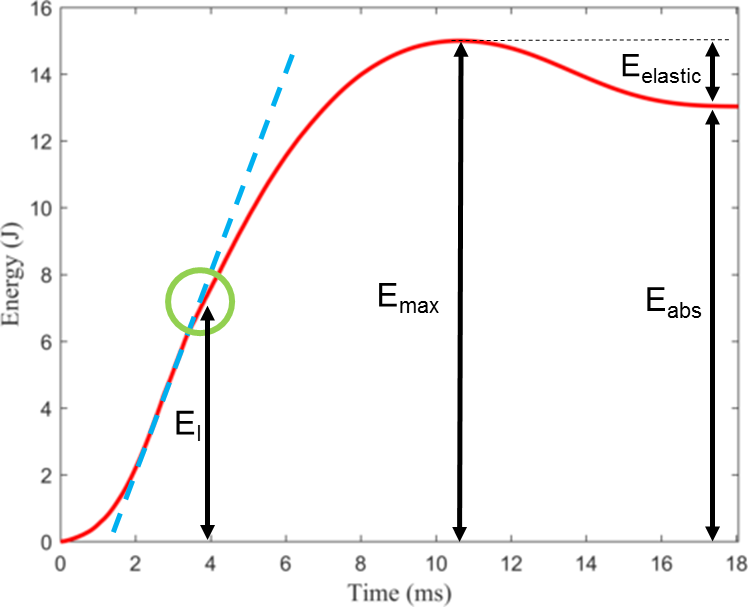}
		\label{1525}
	}
	\subfigure[]{
		\includegraphics[height=5.0cm]{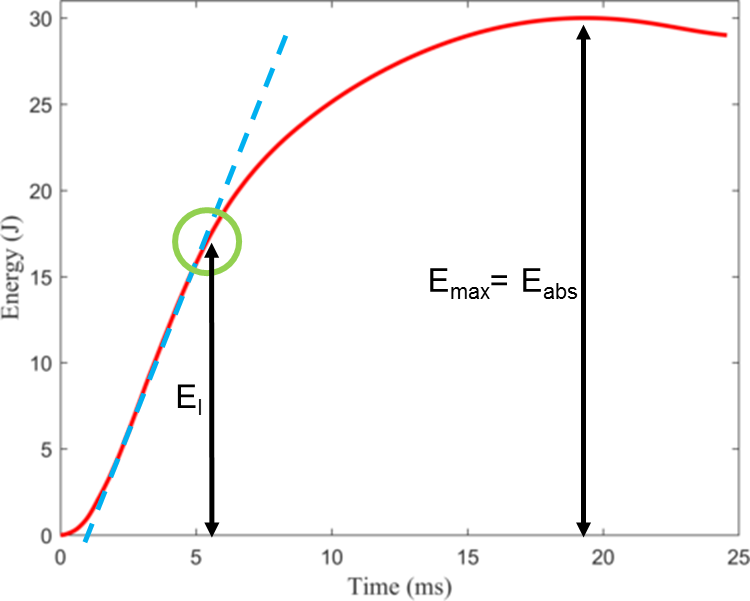}
		\label{3025}
	}
	\subfigure[]{
		\includegraphics[height=5.0cm]{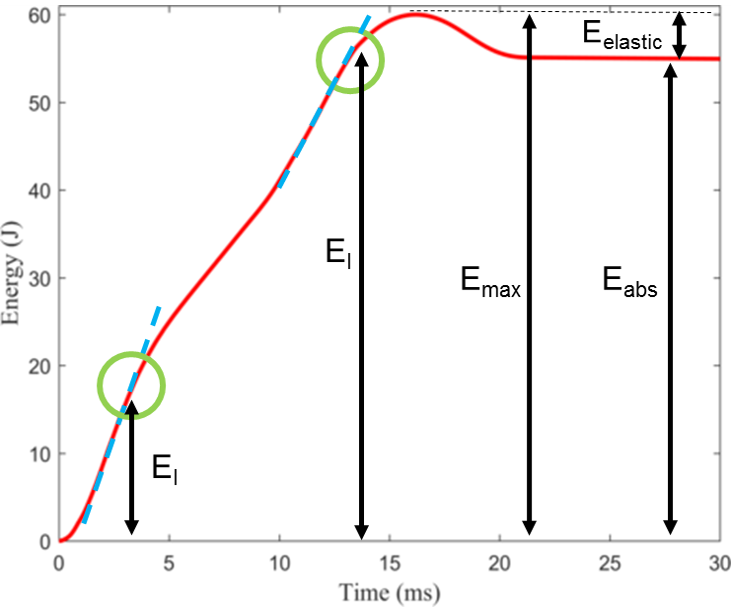}
		\label{6025}
	}
	
	\caption[Optional caption for list of figures]{Representative energy-time responses for samples impacted at 25 $^{\circ}$C at:(a)7.5 J, (b) 15 J, (c) 30 J and (d) 60 J.}
\end{figure}

In addition to the contact force-displacement graphs, the extent of energy absorption and corresponding damage mechanisms presented next are used to characterize the impact behavior of sandwich composites. Figure~\ref{7525} shows the energy-time response of a sample tested under an impact energy of 7.5 J at 25 $^{\circ}$C. The damage initiation, $E_{I}$, is the point where the maximum contact force occurs, beyond which the top facesheet and core are damaged. The maximum impact energy, $E_{max}$, corresponds to the impacted energy. The post-peak response labeled as $E_{abs}$ (plateau region) corresponds to the energy absorbed by the sample through internal deformation and damage, which in this case (7.5 J at 25 $^{\circ}$C) is predominantly delamination in the top facesheet. The difference between $E_{max}$ and $E_{abs}$ is the elastic energy $E_{elastic}$, which is the energy not absorbed by the specimen and is returned to the system by the rebound of the striker. 

Figures \ref{1525}, \ref{3025} and \ref{6025} show the energy-time response of a sample tested at 15 J, 30 J and 60 J each at 25 $^{\circ}$C. These graphs were obtained from the impact tests performed in this study.

In contrast to the energy-time response of the samples impacted at 7.5 J, the samples impacted at 15 J absorbed more energy and the slope of this curve reduced beyond $E_{I}$. This is attributed to the damage mechanisms such as fiber fracture and core crushing observed at 15 J. In this case, the striker rebounded from the sample which can be observed by the elastic energy, $E_{elastic}$. In Figure \ref{3025} at 30 J, $E_{max} = E_{abs}$, which implies that the sample completely absorbed the impacted energy. The reduced slope beyond $E_{I}$ indicates that the top facesheet was completely perforated and the core was penetrated by the striker. Next, recall that two peak forces were observed in the force-displacement responses at 60 J (Figure \ref{60}) due to damage in the top and bottom facesheets along with the core. Therefore, in Figure \ref{6025} at 60 J, the first regime with reduced slope corresponds to the perforation of the top facesheet and core penetration, while the second reduced slope corresponds to fiber fracture at the bottom facesheet and debonding between the core and the bottom facesheet. A small amount of elastic energy is remaining as a result of the striker rebounding from the bottom facesheet.

\begin{figure}[!htb]
	\centering
	\subfigure[]{
		\includegraphics[height=5.0cm]{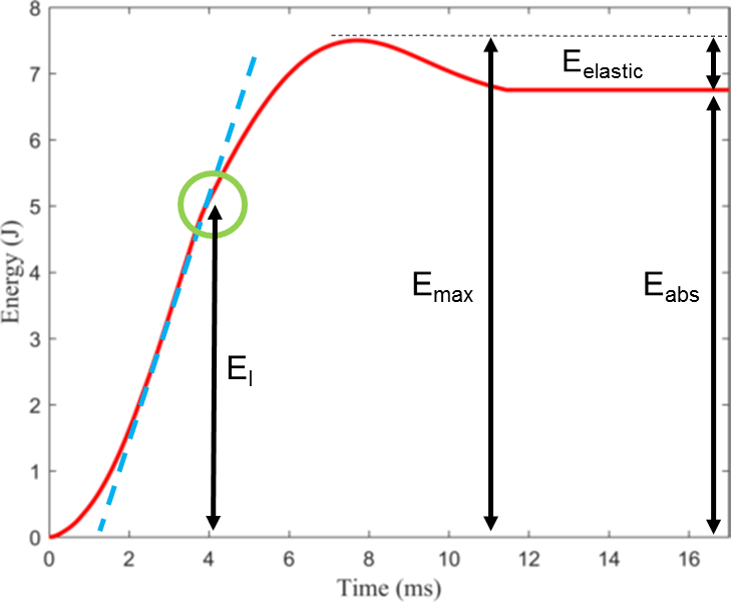}
		\label{7550}
	}
	\subfigure[]{
		\includegraphics[height=5.0cm]{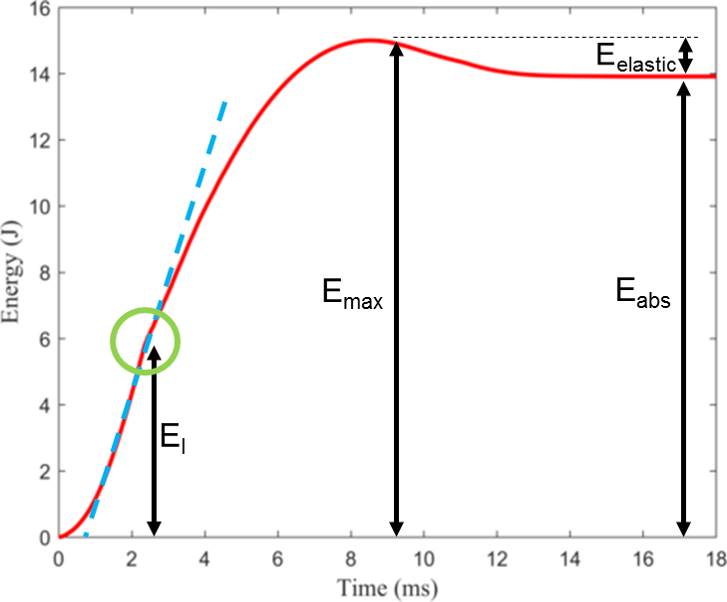}
		\label{1550}
	}
	\subfigure[]{
		\includegraphics[height=5.0cm]{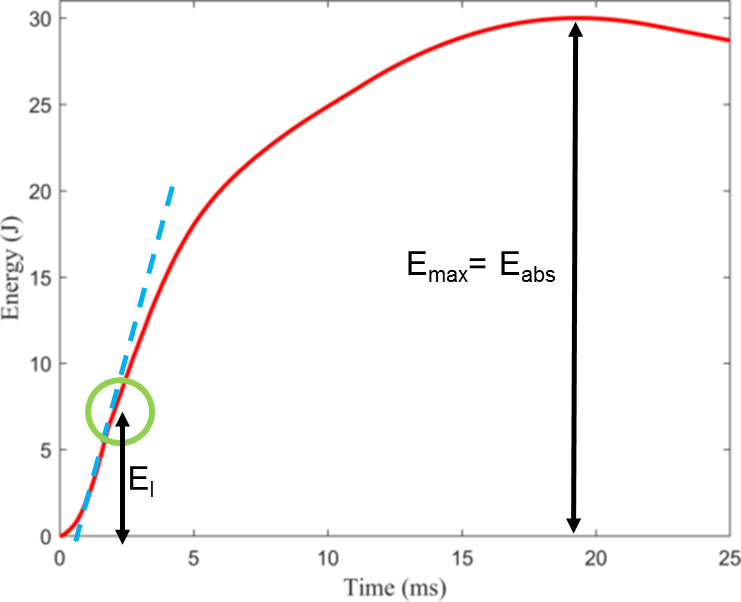}
		\label{3050}
	}
	\subfigure[]{
		\includegraphics[height=5.0cm]{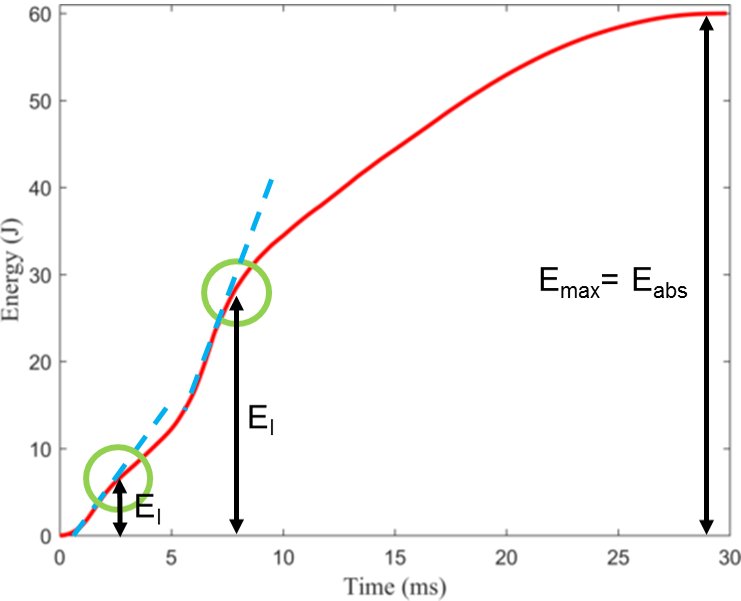}
		\label{6050}
	}
	
	\caption[Optional caption for list of figures]{Representative energy-time responses for samples impacted at -50 $^{\circ}$C at:(a)7.5 J, (b) 15 J, (c) 30 J and (d) 60 J.}
\end{figure}

Energy-time responses of samples tested under impact energies of 7.5 J, 15 J, 30 J and 60 J each at -50 $^{\circ}$C are shown in Figures \ref{7550}, \ref{1550}, \ref{3050} and \ref{6050}, respectively. These graphs were obtained from the impact tests performed in this study. Under 7.5 J, a distinct reduction in the slope of the curve after $E_{I}$ is observed in contrast to the samples at room temperature. This implies that even at 7.5 J of impact energy at -50 $^{\circ}$C, fiber fracture and delamination in the top facesheet and minimal core crushing were observed. Figure~\ref{1550} shows the energy-time response of a sample tested at 15 J at -50 $^{\circ}$C. Again, the slope of the curve reduced beyond $E_{I}$ due to fiber fracture and delamination at the top facesheet and core fracture. At 7.5 J and 15 J, the striker rebounded from the samples, which is manifested as the remaining elastic energy in the plateau regions of the graphs. The energy-time response of a sample tested under 30 J at -50 $^{\circ}$C is shown in Figure ~\ref{3050}. $E_{max}$ = $E_{abs}$ implies that the impacted energy was completely absorbed by the sandwich composite and the striker did not rebound from the specimen. This is similar to that observed at room temperature. The reduced slope beyond $E_{I}$ indicates complete perforation of the top facesheet and core penetration by the striker. 

At 60 J, the sandwich composite samples experienced two peaks in their contact force-displacement responses (Figure \ref{60}) in all in-situ test temperatures including  -50 $^{\circ}$C. Correspondingly, the energy-time response of a sample tested at 60 J at -50 $^{\circ}$C manifested two zones of sudden slope reduction represented by $E_{I}$ as shown in Figure \ref{6050}. The first slope reduction indicates complete perforation of the top facesheet and core penetration by the striker. The second slope reduction corresponds to damage in the bottom facesheet and debonding between the core and bottom facesheet. In contrast to the room temperature case, here $E_{max}$ = $E_{abs}$, which implies that the striker did not rebound as it perforated the bottom facesheet of the sample. It can be concluded that the samples tested at low temperatures (0 $^{\circ}$C, -25 $^{\circ}$C and -50 $^{\circ}$C) experienced more damage than those at 25 $^{\circ}$C. This is manifested by an increase in energy absorption at low temperatures under all the impact energies.

\subsection{Energy profile diagram}
\begin{figure}[!htb]
	\begin{center}
		\includegraphics[height=10.0cm]{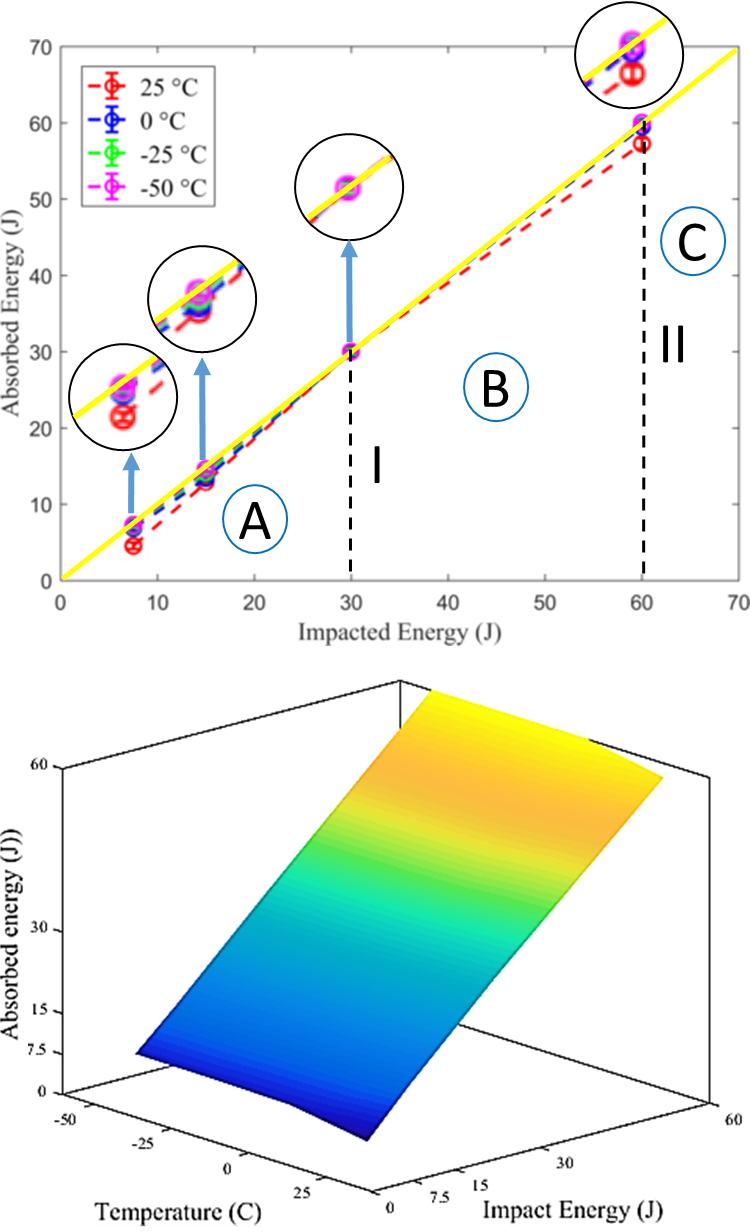}
		\caption{Energy profile diagram for the sandwich composite specimens.}\label{EPD}
	\end{center}
\end{figure}

To better understand the energy absorption process, an energy profile diagram was generated that shows the relationship between the absorbed energy $E_{abs}$ and the impacted energy $E_{max}$ \cite{Atas2010}. The absorbed energy was obtained by integrating the area under the curve of the contact force-displacement responses for each test conducted. Figure~\ref{EPD} shows the energy profile diagram, where each point on the graph represents the average value of the absorbed energy corresponding to an impacted energy at a particular temperature. The yellow line connects the points when the impact energy is equal to the absorbed energy. ``I'' represents the perforation threshold for the top facesheet. Regardless of the in-situ test temperature under 30 J, the striker completely perforated the top facesheet. ``II'' represents the perforation threshold for the bottom facesheet only at low temperatures. That is, the striker only caused minimal fiber breakage at the bottom facesheet at 25 $^{\circ}$C. Hence, the absorbed energy was less than the impacted energy, resulting in the rebound of the striker. At 0 $^{\circ}$C, the samples absorbed almost all the energy due to an increase in debonding between the bottom facesheet and the core. At -25 $^{\circ}$C and -50 $^{\circ}$C, the striker got lodged into the bottom facesheet, resulting in perforation of the bottom facesheet.

In summary, there are three zones in the energy profile diagram: A, B and C. Zone A represents the impact energy where the main damage mechanisms were fiber fracture, delamination and matrix cracking at the top facesheet and core crushing. Zone B represents the impact energy where the main damage mechanisms were perforation of the top facesheet and core fracture. Zone C represents the impact energy where the main damage mechanisms were perforation of the top facesheet, transverse shear fracture in the core, core crushing, debonding between the bottom facesheet and the core, and fiber fracture at the bottom facesheet.

\subsection{Damage mechanisms}
\begin{figure}[h!]
	\centering
	\subfigure[]{
		\includegraphics[height=9.00cm]{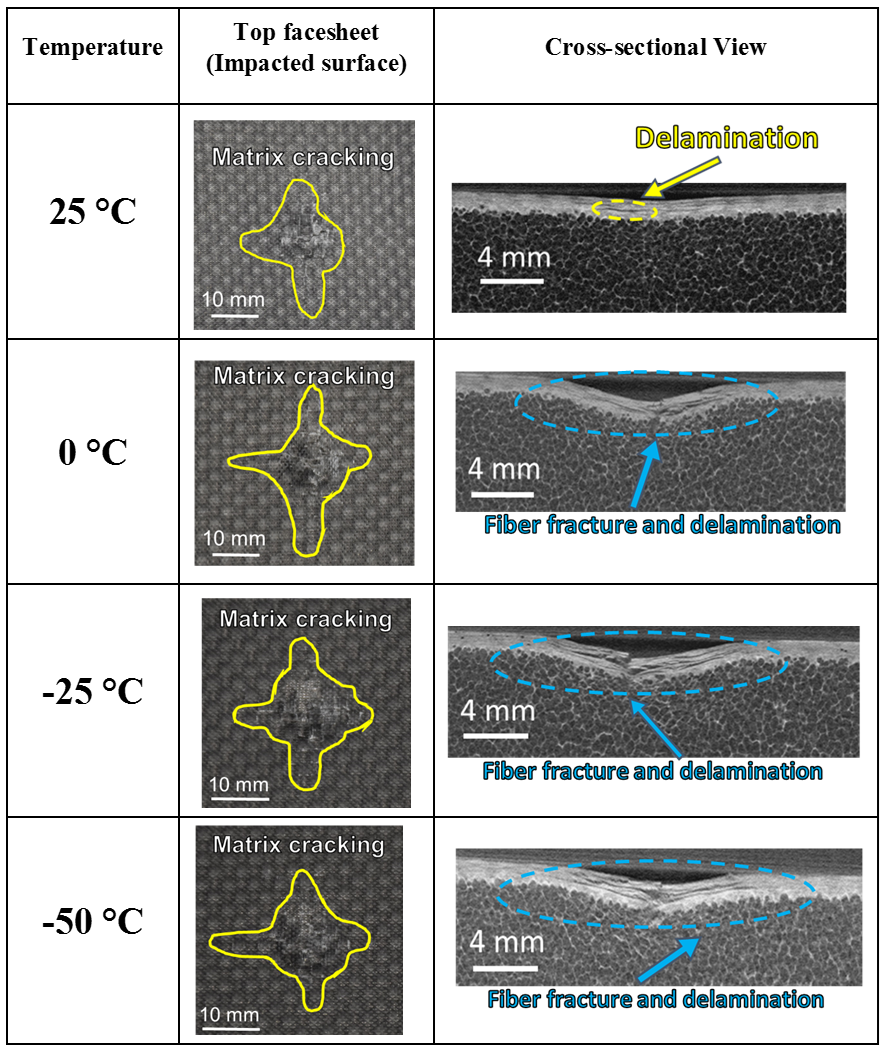}
		\label{temp_75}
	}
	\subfigure[]{
		\includegraphics[height=9.cm]{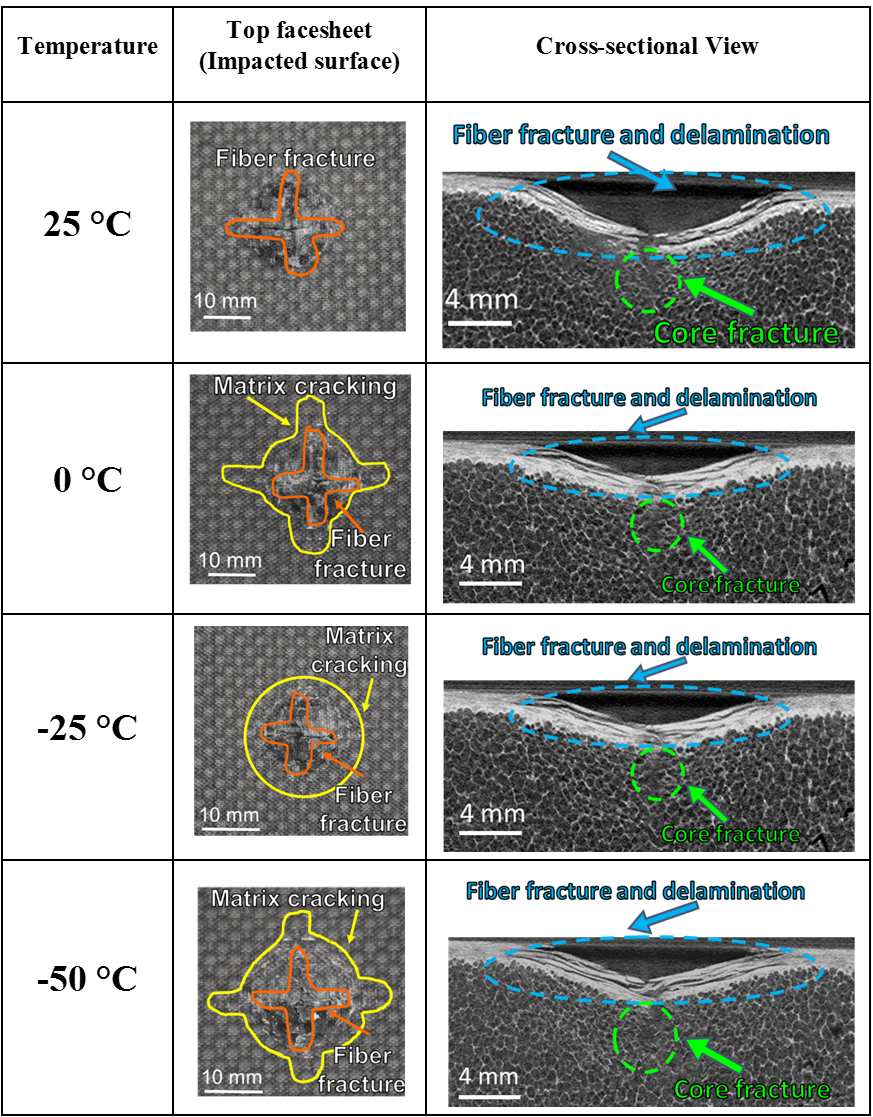}
		\label{temp_15}
	}
	\caption[Optional caption for list of figures]{Top facesheet and cross-sectional view of the sandwich composite specimens impacted at (a) 7.5 J at 25 $^{\circ}$, 0 $^{\circ}$C, -25 $^{\circ}$C and -50 $^{\circ}$C, and (b) 15 J at 25 $^{\circ}$, 0 $^{\circ}$C, -25 $^{\circ}$C and -50 $^{\circ}$C. }
\end{figure}

\begin{figure}[h!]
	\centering
	\subfigure[]{
		\includegraphics[height=9.0cm]{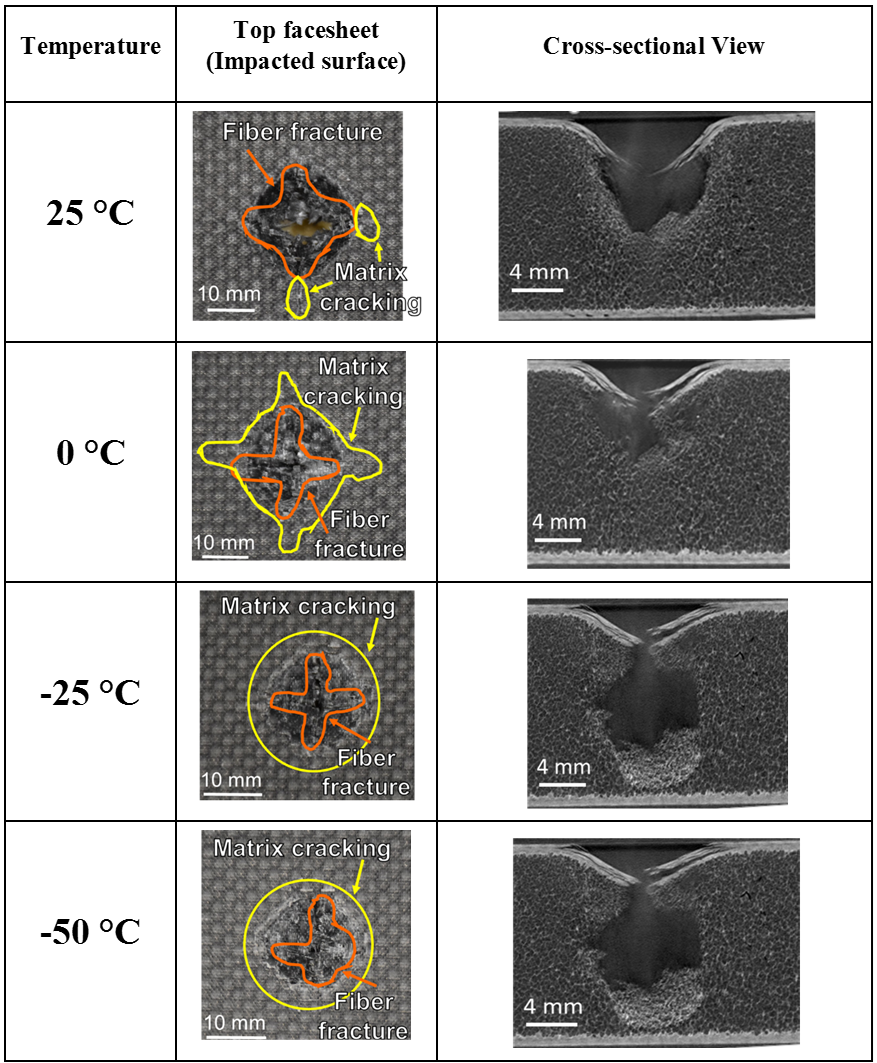}
		\label{temp_30}
	}
	\subfigure[]{
		\includegraphics[width=8.3cm]{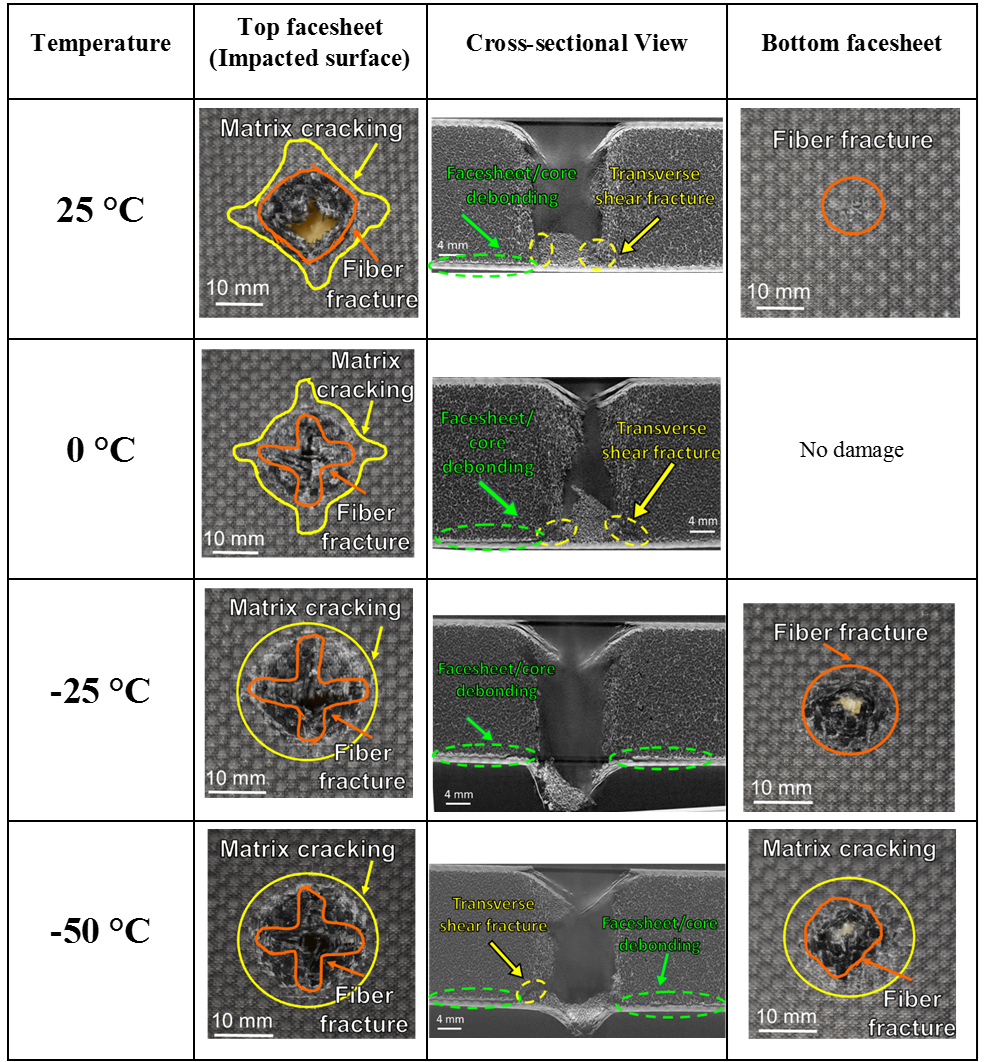}
		\label{temp_60}
	}
	\caption[Optional caption for list of figures]{Top facesheet and cross-sectional view of the sandwich composite specimens impacted at (a) 30 J at 25 $^{\circ}$, 0 $^{\circ}$C, -25 $^{\circ}$C and -50 $^{\circ}$C, and (b) 60 J at 25 $^{\circ}$, 0 $^{\circ}$C, -25 $^{\circ}$C and -50 $^{\circ}$C.}
\end{figure}

Micro-CT scans of the interior regions of samples and optical images of the impacted and back face were obtained to characterize the damage mechanisms in the sandwich composite samples after testing. Figures~\ref{temp_75}, ~\ref{temp_15}, ~\ref{temp_30} and ~\ref{temp_60} show the impacted face of the top facesheet, back face  of the bottom facesheet and cross-sectional view for the samples impacted at 7.5 J, 15 J, 30 J and 60 J, respectively, at all temperatures. The only impact energy that damaged the bottom facesheet was 60 J. Therefore, the images of the bottom facesheet of samples impacted at 7.5 J, 15 J and 30 J are not shown. The regions with different damage mechanisms are highlighted as matrix cracking, fiber fracture, delamination, core fracture, transverse shear fracture and facesheet/core debonding in these images. 

The samples impacted under 7.5 J of energy at 25 $^{\circ}$C experienced delamination and matrix cracking at the top facesheet as shown in Fig.~\ref{temp_75}. At 0 $^{\circ}$C, -25 $^{\circ}$C and -50 $^{\circ}$C, the dominant damage mechanisms were matrix cracks, fiber fracture and delamination at the top facesheet while the core experienced small localized compression. The bottom facesheet was not damaged. The fiber fracture and delamination are manifested as a sudden drop in load in the contact force-displacement responses. The samples impacted under 15 J of energy at all temperatures experienced fiber fracture, matrix cracking and delamination at the top facesheet and localized core crushing/fracture as shown in Figure~\ref{temp_15}. Again, the bottom facesheet did not exhibit any damage. The samples impacted under 30 J of energy at all temperatures exhibited penetration of the striker into the core as shown in Figure~\ref{temp_30}. The top facesheet was perforated with evident matrix cracking surrounding this area. As the temperature decreased, the penetration of the striker into the core increased. This is attributed to the brittleness of the top facesheet and the matrix at low temperatures, which enabled the striker to perforate the top facesheet easily and to cause more damage to the core. There was no evident damage at the bottom facesheet at 30 J.
The samples impacted under 60 J of energy at all temperatures exhibited varying extent of damage at the back facesheet, with almost complete perforation of the bottom facesheet at -25 $^{\circ}$C and -50 $^{\circ}$C as displayed in Figure~\ref{temp_60}. At 25 $^{\circ}$C, the top facesheet was completely perforated, the foam core was crushed up to the bottom facesheet. This resulted in debonding between the facesheet/core and transverse shear fracture in the core. With reducing temperatures, the laminates became stiffer resulting in less deflection and bending. Therefore, at 0 $^{\circ}$C, there was not visible damage at the bottom facesheet. The main failure mechanisms were debonding between the core and the bottom facesheet, transverse shear fracture of the core, and perforation of the top facesheet. At -25 $^{\circ}$C and -50 $^{\circ}$C, the sandwich composite samples were almost completely perforated by the striker, accompanied by the striker getting lodge into the back facesheet. At -50 $^{\circ}$C, it is expected that the sample became more brittle resulting in higher debonding between the bottom facesheet and the core than those tested at -25 $^{\circ}$C.
\section{Conclusion}
Dynamic impact behavior of woven carbon/vinyl ester composites at 25 $^{\circ}$C, 0 $^{\circ}$C, -25 $^{\circ}$C and -50 $^{\circ}$C were investigated in this paper in view of increasing interest in Arctic explorations and the need to characterize these sandwich composites for future arctic applications. Four different impact energies of 7.5 J, 15 J, 30 J, and 60 J were considered for dynamic impact testing at these temperatures. Key observations regarding the contact force, absorbed energy and damage mechanisms were reported and discussed in this paper. Key conclusions are summarized as follows:
\begin{enumerate}
	\item Sandwich composites were rendered stiff and brittle at low temperatures (0 $^{\circ}$C, -25 $^{\circ}$C and -50 $^{\circ}$C) as compared to room temperature (25 $^{\circ}$C). The average bending stiffness values increased with reducing temperatures.
	
	\item For impact energies of 15 J, 30 J and 60 J, the peak contact forces reduced as the temperature decreased. This is attributed to facesheet failure due to the increase of brittleness in the sandwich composites at low temperatures. However, the peak contact forces oscillated at an impact energy of 7.5 J as the temperature decreased. This is attributed to the minimal damage imparted to the facesheets at this impact energy.
	
	\item Damage mechanisms contributed significantly to the amount of energy absorbed at low temperatures. That is, higher degree of damage manifested as the temperature decreased. The corresponding damage modes were more pronounced with increasing impact energies. 
	
	\item The influence of low temperatures on PVC foam core carbon/vinyl ester sandwich composites at lower impact energies (7.5 J) is not as significant as at higher impact energies (15 J, 30 J and 60 J).
	
\end{enumerate}
In conclusion, temperature has a significant influence on the dynamic impact behavior of sandwich composites and can further have a detrimental effect on the residual strengths and durability of composite structures. Further studies to elucidate the influence of other sandwich composite parameters, including composite thickness, varying facesheet/core bonding, stiffness of the core, etc. with varying temperatures and impact energies need to be conducted for reliably using these sandwich composites in low temperatures like the arctic.
\section*{Acknowledgements}
The authors would like to acknowledge the partial support by U.S. Department of Defense (DoD) Office of Naval Research (ONR) Young Investigator Program Award [N00014-19-1-2206] and HBCU/MI Basic Research Grant [W911NF-15-1-0430] for conducting the research presented in this paper.

{\footnotesize
\setcitestyle{square}
	\bibliographystyle{unsrtnat}
	\bibliography{references}
}

\end{document}